\documentclass[showpacs,floatfix,aps]{revtex4}
\usepackage{amssymb,amsmath}
\usepackage[T1]{fontenc}
\usepackage[dvips]{graphics,color}
\usepackage{epsfig}
\usepackage{float}
\newcommand{\bea}{\begin{array}}
\newcommand{\ear}{\end{array}}
\newcommand{\bege}{\begin{equation}}
\newcommand{\enge}{\end{equation}}

\newcommand{\beq}{\begin{eqnarray}}\newcommand{\benu}{\begin{enumerate}}\newcommand{\enu}{\end{enumerate}}
\newcommand{\eeq}{\end{eqnarray}}

\newcommand{\noi}{\noindent}

\definecolor{Magenta}{named}{Magenta}
\definecolor{MidnightBlue}{named}{MidnightBlue}
\definecolor{Emerald}{named}{Emerald}
\definecolor{YellowOrange}{named}{YellowOrange}
\definecolor{Plum}{named}{Plum}

\begin{document}

\title{Gravity with extra dimensions and dark matter interpretation:\\ Phenomenological example via Miyamoto-Nagai galaxy}

\author{P. S. Letelier\footnote{P. S. Letelier passed away in June, 9th 2011. The present
constitutes a posthumous publication.}} \affiliation{Departamento de
Matem\'atica Aplicada, Instituto de Matem\'atica, Estat\'\i stica e
Computa\c c\~ao Cient\'\i fica, Universidade Estadual de Campinas,
Unicamp, 13083-970, Campinas, SP, Brazil.}

\author{C. H. Coimbra-Ara\'ujo}
\email{carlos.coimbra@ufpr.br} \affiliation{Campus Palotina,
Universidade Federal do Paraná, Rua Pioneiro, 2153, 85950-000,
Palotina, Brazil.}


\begin{abstract}
A configuration whose density profile coincides with the Newtonian
potential for spiral galaxies is constructed from a $4D$ isotropic
metric plus extra dimensional components. A Miyamoto-Nagai ansatz is
used to solve Einstein equations. The stable rotation curves of such
system are computed and, without fitting techniques, we recover with
accuracy the observational data for flat or not asymptotically flat
galaxy rotation curves. The density profiles are reconstructed and
compared to that obtained from the Newtonian potential.
\end{abstract}
\maketitle

\section{Introduction}
There is strong observational evidence, primarily from dynamical and
lensing effects, that galactic disks, cluster of galaxies, and a
smoothly distributed cosmological background point to the existence
of the so called ``dark matter''. In galactic disks, where Newtonian
gravitational theory would have been expected to be an excellent
description, accelerations of stars and gas, as estimated from
Doppler velocities are much larger than those due to the Newtonian
field generated by the visible matter in those systems (the plateau
anomaly in rotation curves of galaxies) \cite{oort}. Rotation curves
are the major tool for determining the distribution of mass in
spiral galaxies, and are also important to study kinematics and to
infer the evolutionary histories in galactic systems. Historically
they are the most basic and classic manner to infer the presence of
dark matter in galaxies (for a complete review about rotation
curves, see for instance \cite{sofuerubin}). On the other hand, it
is verified that cluster of galaxies are composed of three main
components: $\sim 5\%$ in mass is the optically luminous baryonic
matter in hundred of bright galaxies; $\sim 15\%$ is in the form of
a bright X-ray inter-cluster gas; and the remaining $\sim 80\%$ is
some sort of non-baryonic ``missing mass''. Such techniques plus the
temperature fluctuations in the Cosmic Background Radiation have
been regarded as confirming the dark matter existence.


In \cite{coimbra1}, a thin disk constructed from a space-time
endowed with extra dimensions just provide the needed extra
parameters to construct, without dark matter, a configuration that
mimics a generic and idealized axially symmetric galaxy. At this
same reference, an impressive result was also obtained for
gravitational lensing effects for a spherical cluster living within
the same space-time. These outcomes are directly related to the dark
matter problem, as it is explained above, in the sense that ``dark
matter'' can be translated as an ignorance to explain why spiral
galaxies have exotic rotation curves (see e.g.
\cite{sofuerubin,rotation} and references therein), galaxy clusters
have a greater amount of ray deflection than expected for
gravitational lensing \cite{fort} (and also mass when confronted to
virial theorem \cite{oort}), and the universe appears to have an
unexpected few fraction of baryons produced during nucleosynthesis
\cite{steigman}.

In the present text it is given a phenomenological example where the
previous thin disk model is naturally extended by using an isotropic
configuration to construct a galaxy as mentioned in
\cite{isotropic}. Unlike the results obtained for the thin disk
model, where the galaxy is interpreted as a flux and counterflux of
geodesic particles, at the present article we obtain a stable
configuration similar to a spiral galaxy, i.e., where we have a
central bulge and a thick disk displayed in a particular density
profile. In this way, the ``artificial'' galaxy constructed by the
previous thin disk is substituted here by a richer model and a more
realistic configuration. The design is totaly carried out by a $4D$
isotropic configuration living in a multidimensional universe. It is
constructed a $6D$ galaxy following the arguments about the
simplicity of self-gravitating objects living in six dimensions
(even number of space-time dimensions) presented in \cite{coimbra1}.
The obtained Einstein equations allow the system to be solved by
Miyamoto-Nagai solutions. Such solutions are important because they
describe a family of self-gravitating configurations that can be
seen as three-dimensional models for the distribution of mass in
galaxies \cite{isotropic,vogt}.


The present work is organized as follows: in Section \ref{sec:field}
the field equations are calculated from a $4D$ isotropic metric plus
extra terms and a Miyamoto-Nagai ansatz is used to solve them. In
Section \ref{sec:rotation} a general equation for the circular
geodesics in the planar part of the configuration is calculated and
it is argued that such equation could approximately represent the
rotation curves and in Section \ref{sec:stability} we use a general
relativistic exact method to calculate the stability of such curves.
Many spiral galaxies can be constructed from our results and in
Section \ref{sec:ngc} it is presented an example for a well known
galaxy (NGC 3198) and the density profile from the obtairned
rotation curves is recovered and compared with the Newtonian
potential and other examples. Finally in Section
\ref{sec:conclusions} we present some concluding remarks. In what
follows we use $c=1$ and $G=1$ (and do not consider possible
variations of $c$ or $G$ with space, time or number of dimensions).

\section{Field equations}\label{sec:field}

Consider a generalization where our universe has $D = 4 + n$
dimensions. For an Einstein--Hilbert gravitational action we have

\begin{equation}
S=\frac{1}{16\pi}\int \mathrm{d}^4x\mathrm{d}^n y
\sqrt{-^{(4+n)}g}~^{(4+n)}R,
\end{equation}

\noi what leads to the field equations \bege\label{einstein1}
~^{(4+n)}G_{AB}=-8\pi ~^{(4+n)}T_{AB}, \enge \noi where $A,B = 0, 1,
..., 4+n-1$, $y$ are the extradimensions and the indices $(4+n)$
tells about the multidimensional nature of the action. In many
theories of compactified extra dimensions it is calculated a new
Newton constant $G$ for dimensions greater than $1+3$. Here we relax
about compactification and maintain the usual value for $G$. First
of all, let us consider the case of axial-symmetric 4D space-times
whose metric can be written in a isotropic form in cylindrical
coordinates $(t, R, z, \varphi)$: \beq\label{general_metrica}
\mathrm{d}s^2 = \mathrm{e}^{\nu(R,z)}\mathrm{d}t^2 -
\mathrm{e}^{\lambda(R,z)}(\mathrm{d}R^2 + \mathrm{d}z^2 +
R^2\mathrm{d}\varphi^2). \eeq

A general relativistic formulation for the Newtonian well known galaxy models can be written in the form of the Schwarzschild metric in isotropic coordinates (see e.g. \cite{vogt}). Also, as a form to extend the formalism developed in \cite{coimbra1}, we introduce $n$ extradimensional coordinates. For this case, it is
showed in \cite{coimbra1} that non-exotic matter is only possible
(including the fact about a satisfactory Huygens principle) if $n$
is even. Let us consider e.g. the simplest case, where $n=2$:

\beq\label{metrica}
\mathrm{d}s^2=&&\frac{(1-f)^2}{(1+f)^2}\mathrm{d}t^2-(1+f)^4[\mathrm{d}R^2+\mathrm{d}z^2+R^2\mathrm{d}\varphi^2]\nonumber\\
&&-\mathrm{e}^{-k}\mathrm{d}x^2-\mathrm{e}^{k}\mathrm{d}y^2, \eeq
\noi where $f=f(R,z)$ and $k=k(R,z)$. The field equations
(\ref{einstein1}) yield the next expressions for the components of
the energy-momentum tensor. The $T^t_t$ component is calculated as:

\bege T^t_t=\frac{1}{32\pi(1+f)^7}\left[16\left(f_{,RR}+f_{,zz}+\frac{f_{,R}}{R}\right)(1+f)^2+(k_{,R}^2-k_{,z}^2)(f^3+f^2+f+1)\right].\enge

\noi Here as a first approximation it will be assumed that the density profile coincides with the Newtonian potential in $3D$ and so one can rewrite the above expression only with visible components as

\bege \label{eq_Ttt}
 T^t_t=\frac{1}{2\pi (1+f)^5}\left(f_{,RR}+f_{,zz}+\frac{f_{,R}}{R}\right),
\enge \noi and therefore the constraint

\bege k_{,R}^2-k_{,z}^2=0. \enge

\noi It is a very interesting approach because we can at the same time obtain a observational quantity that can be compared with what is observed in real galaxies and a differential equation that gives the aspect of the extradimensional function $k(R,z)$. The general solutions for this last have the form \bege
\label{eq_k} k=k_1(z-R)+k_2 \;\; \mathrm{or}\;\;
k=k_1(z+R)+k_2,\enge \noi where $k_1$ and $k_2$ are constants, and
for simplicity we will consider $k_2=0$. Also, with a such
constraint, the $4D$ part of pressure solutions are the same as
obtained by Vogt and Letelier \cite{vogt}: \bege T^R_R
=\frac{1}{4\pi(1+f)^5(1-f)} \left(
ff_{,zz}+\frac{ff_{,R}}{R}+2f_{,R}^2-f_{,z}^2 \right) \mbox{,}
\label{eq_TRR} \enge \bege T^z_z =\frac{1}{4\pi  (1+f)^5(1-f)}
\left( ff_{,RR}+\frac{ff_{,R}}{R}+2f_{,z}^2-f_{,R}^2 \right)
\mbox{,} \label{eq_Tzz}\enge \bege T^R_z =T^z_R= -\frac{1}{4\pi
(1+f)^5(1-f)} \left( ff_{,Rz}-3f_{,R}f_{,z} \right) \mbox{,}
\label{eq_TRz}\enge \bege T^{\varphi}_{\varphi} = \frac{1}{4\pi
(1+f)^5(1-f)} \left[ f \left( f_{,RR}+f_{,zz} \right)
-f_{,R}^2-f_{,z}^2 \right] \mbox{.} \label{eq_Tfifi} \enge \noi The
extradimensional pressure part has the form \bege
T^x_x=\frac{e^-k}{4\pi (1+f)^5}
\left(f_{,RR}+f_{,zz}+\frac{f_{,R}}{R}\right), \enge \bege T^y_y =
-T^x_x.\enge \noi The last two equations permit one to interpret the
model as part of some universal extradimensional theory. The energy
density is given by $\rho=T^t_t$, and the stresses (pressures or
tensions) along a particular direction read $P_i=-T^i_i$ when the
energy-momentum tensor is diagonal. It is a surprising fact that the
component $T^t_t$ is proporcional to the usual Laplacian of the
function $f$ in $flat$ cylindrical coordinates. Note that in the
Newtonian limit when $f \ll 1$, Eq.\ (\ref{eq_Ttt}) reduces to
Poisson equation
\begin{equation} \label{eq_poisson}
\nabla^2 \Phi = 4\pi \rho_N \mbox{,}
\end{equation}
if the function $f$ is related to the gravitational potential $\Phi$
by
\begin{equation} \label{eq_f}
f=-\frac{\Phi}{2} \mbox{.}
\end{equation}
In this case, $\rho \rightarrow \rho_N$ and the energy conditions
for the disk has no exotic matter
\begin{equation}
\rho + \sum_i P_i>0.
\end{equation}
The energy-momentum tensor will be diagonal ($T^R_z=T^z_R=0$)
provided $f$ has the form \bege f=\frac{C}{\sqrt{w(R)+g(z)}}, \enge
where $C$ is a constant and $w(R)$ and $g(z)$ are arbitrary
functions. Complementarily, $T^R_R$ and $T^\varphi_\varphi$ will be
equal (isotropic radial and azimuthal stresses) only if $w(R)=R^2$.
The density profile $\rho$ can now be deduced, e.g., by
Miyamoto-Nagai solutions \cite{isotropic}, which represent
stratifications of mass in the central bulges and in the disk parts
of galaxies. In this case, the simplest gravitational potential that
provides diagonal components is \bege\label{miyamoto}
\Phi(R,z)=-\frac{M}{\sqrt{R^2+(a+\sqrt{z^2+b^2})^2}}, \enge \noi
where $a, b$ are positive constants.

\begin{figure}
\centering
\includegraphics[width=9cm]{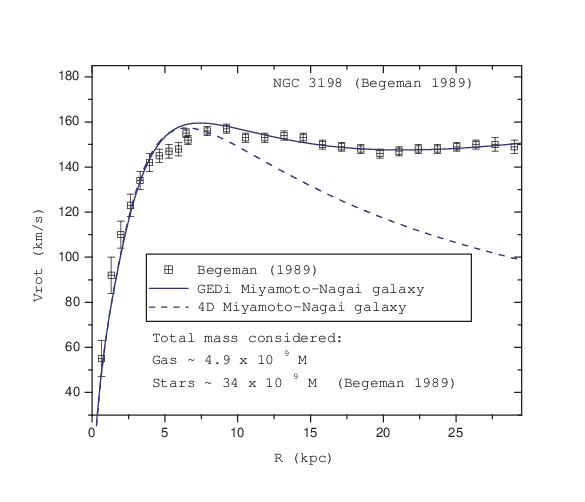}
\caption{\small Rotation curves of NGC3198 implemented from
gravitation with extra dimensions (GEDi, here with six dimensions)
and a pure $4D$ Miyamoto-Nagai solution. The plots come only from
the mass due to gas and stars, i.e., no dark matter is considered.
The discrepancy at 5-7 kpc is due to the fact that our method is not
a fitting technique and also the gas dynamics was not modeled. The
result comes from circular geodesics of test particles around the
designed configuration. We use $b/a=0.01$ (that models a disky
galaxy) and the stability parameter used here is $k_1=10^{-6}$. To
address stability we use also (for extradimensional parameters)
$C_x=0.2$ and $C_y=0.8$.} \label{fig1}
\end{figure}

The corresponding three-dimensional density derived from Eq.
(\ref{eq_poisson}) is \bege \label{eq_d}
\rho_N(R,z)=\frac{b^2M}{4\pi}\frac{aR^2+(a+3\sqrt{z^2+b^2})(a+\sqrt{z^2+b^2})^2}{[R^2+(a+\sqrt{z^2+b^2})^2]^{5/2}(z^2+b^2)^{3/2})},\enge
\noi and now the function $f(R,z)$, according to Eq. (\ref{eq_f}),
is \bege\label{eq_f2}
f(R,z)=\frac{M}{2\sqrt{R^2+(a+\sqrt{z^2+b^2})^2}}. \enge

\noi As we will see in next sections, the gravitational potential
only can be approximated by $f(R,z)$ in the Newtonian limit. By now,
the true gravitational potential will be calculated from the
circular velocity of some test particle in the system.


\section{Approximated rotation curves from circular
geodesics}\label{sec:rotation} Now, the particles of such
configuration describe trajectories that can be derived by
calculating the system geodesic equations. Usually, when there are
some stresses (and there are, both radial and azimuthal), the
configuration can hardly be interpreted in terms of particles moving
on circular geodesics. However, the assumption of geodesic motion is
only valid for the case of a particle moving in a very diluted gas
like the gas made of stars that models a galaxy disk. Given this
important statement, it is possible to obtain the tangential
velocities of the disk particles (i.e. the approximated planar
rotation curves) from geodesic equations. Assuming $\dot{R}=0$ and
$\dot{z}=0$ (the particles have no radial motion and for simplicity
are confined in the surface $z=0$), the metric (\ref{metrica}) can
be rewritten as \bege \frac{(1-f)^2}{(1+f)^2}\dot{t}^2-(1+f)^4
R^2\dot{\varphi^2}-\mathrm{e}^{-k}\dot{x}^2-\mathrm{e}^{k}\dot{y}^2=1
\enge \noindent where $\dot{x}^A=\mathrm{d}x^A/\mathrm{d}s$, which
gives \bege \label{lagrange}
\dot{t}^2=\left(\frac{1+f}{1-f}\right)^2[1+(1+f)^4
R^2\dot{\varphi^2}+\mathrm{e}^{-k}\dot{x}^2+\mathrm{e}^{k}\dot{y}^2].\enge
\noi The Euler-Lagrange equations for $x$ and $y$ coordinates permit
to calculate the following geodesic equations \bege \label{xdot}
(\mathrm{e}^{-k} \dot{x})^\cdot=0 ; \;\;\; \mathrm{e}^{-k} \dot{x} =
C_x, \enge \bege\label{ydot} (\mathrm{e}^k \dot{y})^\cdot=0 ;\;\;\;
\mathrm{e}^k \dot{y} = C_y, \enge \noi where $C_x$ and $C_y$ are
integration constants. We can fix the values of $C_x$ and $C_y$ by
those calculated in \cite{coimbra2}, where it is obtained a stable
planar configuration.

Another equation is obtained by derivating (\ref{lagrange}) in $r$
and using eqs. (\ref{xdot}) and (\ref{ydot}) \begin{eqnarray}
\label{lagrange2}
&&2f_{,R}\left[\frac{(1-f)}{(1+f)^2}+\frac{(1-f)^2}{(1+f)^3}\right]\dot{t}^2
\\\nonumber&&+ 2R(1+f)^3[(1+f)+2R f_{,R}]\dot{\varphi}^2 -
k_{,R}(\mathrm{e}^{-k} \dot{x}^2 - \mathrm{e}^{k} \dot{y}^2) = 0.
\end{eqnarray} \noi Eqs. (\ref{lagrange}) and (\ref{lagrange2}) form a system
of equations whose variables are $\dot{\varphi}^2$ and $\dot{t}^2$.
By solving the system, it is possible to calculate the rotation
curves $V_C$ by
\begin{equation}\label{eq_rot}
V_C=\sqrt{-\frac{g_{\varphi
\varphi}}{g_{tt}}}\frac{\mathrm{d}\varphi}{\mathrm{d}t}=\sqrt{-\frac{g_{\varphi
\varphi}}{g_{tt}}\frac{\dot{\varphi}^2}{\dot{t}^2}},
\end{equation}

\section{Stability}\label{sec:stability}

\begin{figure}
\centering
\includegraphics[width=9cm]{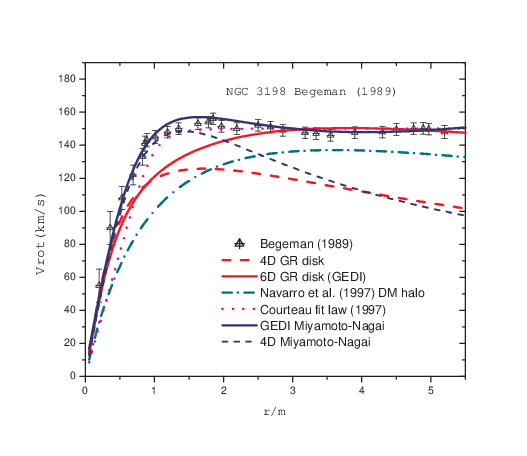}
\caption{\small Rotation curves of NGC3198. Also is considered a
comparison between our model (a gravitation with extra dimensions,
i.e. the GEDi Miyamoto-Nagai), a GEDi thin disk \cite{coimbra1} and
other models or fittings \cite{begeman,navarro,courteau}. The
coordinates are the rotation velocity and the radius per mass.}
\label{fig2}
\end{figure}

For such planar rotation curves, it is needed to calculate the
stability. The most complete method is described in \cite{coimbra2}.
In this case we compute the infinitesimal perturbations in the
geodesic. The perturbation of the geodesic equation
$\ddot{x}^A+\Gamma^A_{BC}\dot{x}^B\dot{x}^C=0$ is done performing
the transformation $x^A\rightarrow x^A+\Delta^A$ --- where
$\Delta^A=(\delta t, \delta r, \delta \varphi, \delta z, \delta x,
\delta y)$ are infinitesimal elements. As showed in \cite{coimbra2},
it is possible to achieve to the following equations for
perturbations:
\begin{equation}\label{perturb}
\ddot{\Delta}^A + 2\Gamma^A_{BC}\dot{x}^B\dot{\Delta}^C +
\Gamma^A_{BC,D}\Delta^D\dot{x}^B\dot{x}^C=0,
\end{equation}
\noindent where $\Gamma^A_{BC}$ are the Christoffel symbols and
$\dot{x}^A$ are proper time derivatives $\mathrm{d}x^A/\mathrm{d}s$
and can be written for a circular orbital motion as
\begin{equation}\label{veloc}
\dot{x}^A=(u^t,0,0,u^t\Omega,u^t C_x,u^t C_y),
\end{equation}
\noindent where $\Omega=V_C/R$, Eq. (\ref{eq_rot}). Assuming
oscillations in all directions with no vertical or extradimensional
restrictions, we get
\begin{equation}
\Delta^{A}=(\delta t, \delta R,\delta z,\delta \varphi,\delta
x,\delta y).
\end{equation}
The non-null computed Christoffel symbols are:
$\Gamma^t_{tR}=\Gamma^t_{Rt}$, $\Gamma^t_{tz}=\Gamma^t_{zt}$,
$\Gamma^R_{tt}$, $\Gamma^R_{RR}$, $\Gamma^R_{Rz}=\Gamma^R_{zR}$,
$\Gamma^R_{zz}$, $\Gamma^R_{\varphi \varphi}$, $\Gamma^R_{xx}$,
$\Gamma^R_{yy}$, $\Gamma^z_{tt}$, $\Gamma^z_{RR}$,
$\Gamma^z_{Rz}=\Gamma^z_{zR}$, $\Gamma^z_{zz}$, $\Gamma^z_{\varphi
\varphi}$, $\Gamma^z_{xx}$, $\Gamma^z_{yy}$, $\Gamma^\varphi_{R
\varphi} = \Gamma^\varphi_{\varphi R}$, $\Gamma^\varphi_{z \varphi}=
\Gamma^\varphi_{\varphi z}$, $\Gamma^x_{R x} = \Gamma^x_{x R}$,
$\Gamma^x_{z x} = \Gamma^x_{x z}$, $\Gamma^y_{R y} = \Gamma^y_{y
R}$, $\Gamma^y_{z y} = \Gamma^y_{y z}$. \noindent Let $x^A$ be an
equatorial circular geodesic in a stationary axisymmetric space-time
(\ref{metrica}), i.e., the worldline $x^A=(t, R=\mathrm{const},
\varphi = \mathrm{const} + \Omega t, z=0, x=\mathrm{const},
y=\mathrm{const})$. Substituting the four velocity (\ref{veloc}),
and from the non-null computed Christoffel symbols, the components
of Eq.(\ref{perturb}) and supposing that the solutions for $\delta
t$, $\delta R$, $\delta z$, $\delta \varphi$, $\delta x$ and $\delta
y$ have a form of harmonic oscillations, $\sim e^{iKs}$, with a
common proper angular frequency $K$, we find:
\begin{widetext}
\begin{eqnarray}
&&(\ddot{\delta t}) + 2\Gamma^t_{tR}u^t(\dot{\delta R})+ 2\Gamma^t_{tz}u^t(\dot{\delta z})=0,\label{1}\\
&&(\ddot{\delta R}) + 2\Gamma^R_{tt}u^t(\dot{\delta t}) + 2\Gamma^R_{\varphi \varphi}u^t\Omega(\dot{\delta \varphi})+[(\Gamma^R_{tt,R}+\Gamma^R_{\varphi \varphi,R}\Omega^2 + \Gamma^R_{xx,R}C_x^2 +\Gamma^R_{yy,R}C_y^2)(u^t)^2]\delta R\\\nonumber&&+[(\Gamma^R_{tt,z}+\Gamma^R_{\varphi \varphi,z}\Omega^2 + \Gamma^R_{xx,z}C_x^2 +\Gamma^R_{yy,z}C_y^2)(u^t)^2]\delta z=0,\label{2}\\
&&(\ddot{\delta z})+2\Gamma^z_{tt}u^t(\dot{\delta t}) + 2\Gamma^z_{\varphi \varphi}u^t\Omega(\dot{\delta \varphi})+[(\Gamma^z_{tt,R}+\Gamma^z_{\varphi \varphi,R}\Omega^2 + \Gamma^z_{xx,R}C_x^2 +\Gamma^z_{yy,R}C_y^2)(u^t)^2]\delta R\\\nonumber&&+[(\Gamma^z_{tt,z}+\Gamma^z_{\varphi \varphi,z}\Omega^2 + \Gamma^z_{xx,z}C_x^2 +\Gamma^z_{yy,z}C_y^2)(u^t)^2]\delta z=0,\label{3}\\
&&(\ddot{\delta \varphi})+2\Gamma^\varphi_{\varphi R}\Omega u^t
(\dot{\delta R})+2\Gamma^\varphi_{\varphi z}\Omega u^t(\dot{\delta
z})=0,\label{4}\\
&&(\ddot{\delta x})+2\Gamma^x_{\varphi R}C_x u^t (\dot{\delta
R})+2\Gamma^x_{x z}C_x u^t(\dot{\delta z})=0,\label{5}\\
&&(\ddot{\delta y})+2\Gamma^y_{\varphi R}C_y u^t (\dot{\delta
R})+2\Gamma^y_{y z}C_y u^t(\dot{\delta z})=0.\label{6}
\end{eqnarray}
\end{widetext}

\begin{figure*}
\begin{center}
$\begin{array}{c@{\hspace{0.01in}}c} \multicolumn{1}{l}{\mbox{\bf
(a)}} &
    \multicolumn{1}{l}{\mbox{\bf (b)}} \\ [-0.33cm]
\epsfxsize=3.30in \epsffile{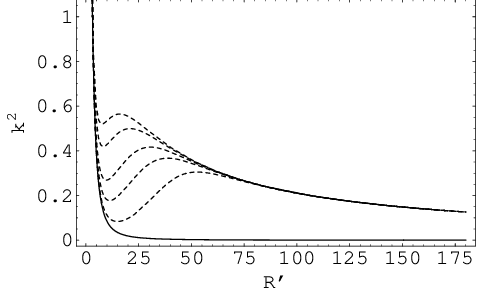} & \epsfxsize=3.40in
    \epsffile{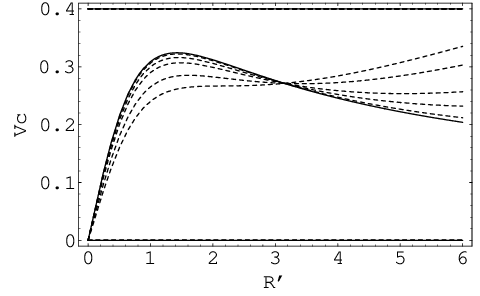} \\ [0.05cm]
\end{array}$
\end{center}
\caption{{\bf (a)} Stability study for the Miyamoto-Nagai
configurations obtained with extradimensions (dotted curves). For
comparison, the full line represents the $4D$ stable curve. Here
$\kappa^2=(K/u^t)^2$>0 represents stable curves. {\bf (b)} The
variation of rotation curves for the range of stable models ploted
in (a).} \label{density3d}
\end{figure*}

The unique real solution for this homogenous system is
\begin{widetext}\begin{eqnarray}\label{epic} K^2=
\frac{R_1}{3}&-&\frac{2^{1/3}(-R_1^2+R_2)}{3\left[2R_1^3-9R_1R_2+27R_3+
\sqrt{4(-R_1^2+3R_2)^3+(2R_1^3-9R_1R_2+27R_3)^2}\right]^{1/3}}\\\nonumber&&+
\frac{1}{3(2^{1/3})}\left[2R_1^3-9R_1R_2+27R_3+\sqrt{4(-R_1^2+3R_2)^3
+(2R_1^3-9R_1R_2+27R_3)^2}\right]^{1/3},\end{eqnarray}
\end{widetext}
\noi where $R_1=(\Gamma^R_{AB,R}+\Gamma^z_{AB,z})u^Au^B$,
$R_2=(\Gamma^R_{AB,R}u^Au^B)(\Gamma^z_{AB,z}u^Au^B)$ and
$R_3=16\Gamma^y_{yR}\Gamma^R_{yy}\Gamma^x_{xz}\Gamma^z_{xx}C_x^2C_y^2(u^t)^4$,
and where $\Gamma^R_{AB,R}u^Au^B=(\Gamma^R_{tt,R}+\Gamma^R_{\varphi
\varphi,R}\Omega^2 + \Gamma^R_{xx,R}C_x^2
+\Gamma^R_{yy,R}C_y^2)(u^t)^2$. The system is stable if the squared
epicyclic frequency $\kappa^2=(K/u^t)^2$ is strictly positive.

This occurs when the signal before the square root in Eq.
(\ref{epic}) is positive and for $k=k_1(z+R)$ in Eq. (\ref{eq_k}).
More than this, the stability test may have an important role to
determinate a stable range of values for the constants $C_x$, $C_y$
and $k_1$ --- this last a constant of integration which came from
the presence of extradimensions for a non-exotic system, Eq.
(\ref{eq_k}). In the case of constants $C_x$ and $C_y$, the values
coincides exactly with the same obtained in \cite{coimbra2}. In the
case of the constant $k_1$, it must be tiny ($k_1 \sim
10^{-5}-10^{-7}$). The usual manner to determine parameters $a$ and
$b$ (that are given in kpc) in Eq. (\ref{eq_f2}) is to obtain them
phenomenologically by observing values of density for the galaxies,
as done in \cite{isotropic}. Usually, the fraction $b/a$ estimates
how disky or spherical is the galaxy. Thus, parameters $a$ and $b$
can be rearranged in the new parameter $b/a$. Usually, disky
galaxies have $b/a\sim 0.01$. Observation of densities plus the
stability study present a complete manner to construct a galaxy for
a system modeled from gravitation with extra dimensions.

\section{Probing the model with real galaxies}\label{sec:ngc}

As showed in \cite{isotropic} one can model a real spiral galaxy by using
a superposition of the central and outbound densities for different values of
$a$ and $b$. Here we will constrain as well the value of $k$ as explained in the
previous section. This is the same as write $\rho = \sum_i \rho(a_i,b_i)$. In other
words, this is the same to write the function $f$, Eq. (\ref{eq_f2}), by
the suporposition of $\sum_i f(a_i,b_i)$. For the case of a spiral galaxy, say, it is
suffice to write for the central bulge and for the disk galaxy part.
It is possible indeed to calculate the rotation curves for many
galaxies (including not only the spiral morphology). As an example,
the spiral galaxy NGC 3198 can be modeled. Considering the surface
density observed on the disk and the morphology characteristics
\cite{begeman} and assuming that the bulge density is not so
different than that observed for Milky Way \cite{isotropic} (where
usually $a$ is set to zero and $b \sim 1$kpc; this means that $b/a
\rightarrow \infty$, or a total spherical $\sim$ 1kpc central
bulge), we obtain a set of stable rotation curves (see Figs.
\ref{fig1} and \ref{fig2}).

The potential presented at Eq. (\ref{miyamoto}) -- i.e. a
Miyamoto-Nagai ansatz -- is purely Newtonian and was useful to
calculate the form of the function $f$ in the metric. Therefore it
is necessary to calculate the true potential related to the problem.
We can translate the potential form by calculating the density by

\bege  \rho=\frac{\nabla^2 \Phi}{4\pi}, \enge

\noi where the potential is reconstructed from the planar circular
geodesics (\ref{eq_rot}) by \bege \Phi = \int^R_0
\frac{V_C^2}{R}\mathrm{d}R, \enge

\begin{figure}
\centering
\includegraphics[width=9cm]{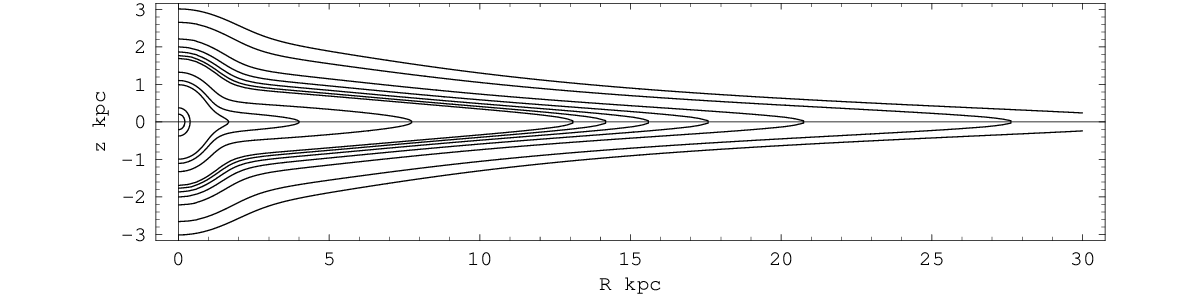}
\caption{The GEDi density contour plots for NGC 3198. We use
$b/a=0.01$ (that models a disky galaxy) and the stability parameter
used here is $k_1=10^{-6}$. To address stability we use also
$C_x=0.2$ and $C_y=0.8$.} \label{density1}
\end{figure}

\noi where $V_C$ is given in Eq. (\ref{eq_rot}). The density contour
plots recovered from such equations is showed in Fig.
\ref{density1}, and our object is very similar to a spiral galaxy.

We can compared such result with other potentials known in literature.
In e.g. \cite{binney}, it is listed some of possible galaxy
potentials. A first one, used here as well (Miyamoto-Nagai
potential-density pair) is compared to the density recovered from
our circular velocities (see Fig. \ref{density3d}).

\begin{figure*}
\begin{center}
$\begin{array}{c@{\hspace{0.01in}}c} \multicolumn{1}{l}{\mbox{\bf
(a)}} &
    \multicolumn{1}{l}{\mbox{\bf (b)}} \\ [-0.33cm]
\epsfxsize=3.30in \epsffile{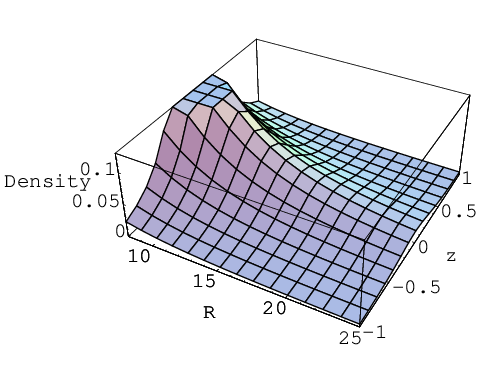} & \epsfxsize=3.40in
    \epsffile{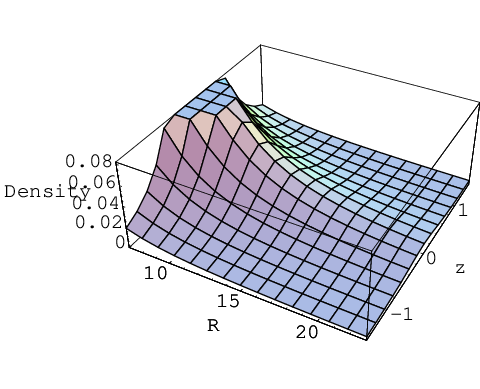} \\ [0.05cm]
\end{array}$
\end{center}
\caption{{\bf (a)} The Miyamoto-Nagai $4D$ model Newtonian density
$\rho_N$ in $M_\odot$pc$^{-3}$ for NGC 3198 (the non-central disk,
where the rotation curve anomaly appears), where $R$ and $z$ are
given in kpc. {\bf (b)} The $3D$ plot of the recovered density
profile for the present model.} \label{density3d}
\end{figure*}

Another interesting potential to be compared is the Satoh's

\bege \Phi^\infty_M(R,z) = -\frac{M}{S},\enge

\bege \rho^\infty_M = \frac{ab^2M}{4\pi S^3(z^2+b^2)}\left[
\frac{3}{a}\left(1-\frac{R^2+z^2}{S^2}\right) + \frac{1}{z^2 + b^2}
\right], \enge

\bege S = [R^2 + z^2 + a(a + 2\sqrt{z^2 + b^2})]^{1/2}.  \enge

\begin{figure}
\centering
\includegraphics[width=9cm]{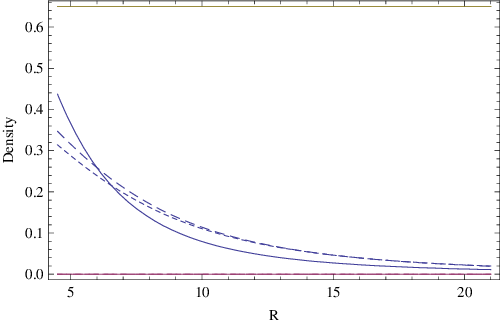}
\caption{Comparison between the density recovered from our circular
velocities (full line), a pure $4D$ Myiamoto-Nagai profile
(long dash line), and a Satoh profile (short dash line). Here the density
is given in $M_\odot$ pc$^{-3}$ and comes from the configuration slice where $z=0$.}
\label{fig:densities}
\end{figure}

Now, for same parameters $a$ and $b$, we plot those densities in
Fig. \ref{fig:densities}. The bit difference between the
reconstructed density profile and conventional Newtonian $4D$
examples, shows that the extra dimensions affect remarkably the
circular velocity profile but, although affect poorly some absolute
values when compared to other potentials (what indeed does not
disagree remarkably with observations), the density contours point
to an object very similar to a spiral galaxy.

\section{Rotation curves are not flat!}

\begin{figure}
\centering
\includegraphics[width=9cm]{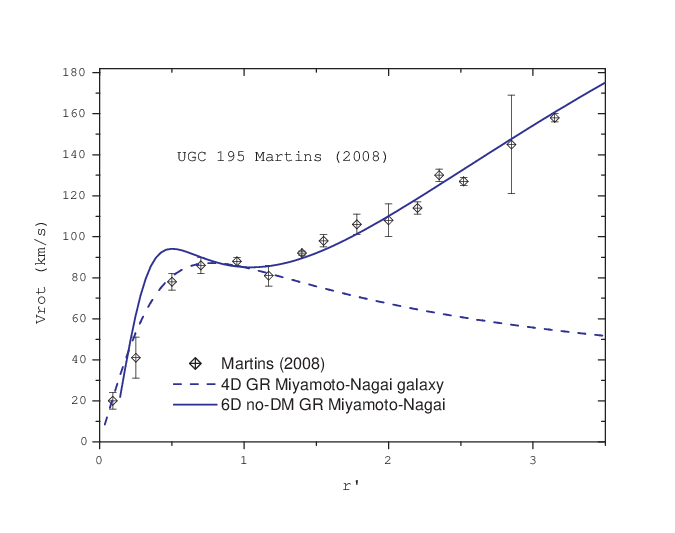}
\caption{The observed non-planar rotation curve of galaxy UGC 195
\cite{martins}, the Miyamoto-Nagai $4D$ expected profile and the
same when it is included more two extra dimensions in computation
(the case of the present model). Parameters used: $b/a=0.01$,
$C_x=0.2$, $C_y=0.9$ and $k_1=10^{-5}$.} \label{fig:nonplanar}
\end{figure}

An essential observational remark is that real galaxy rotation
curves are not flat. Thus the flat rotation curve paradigm should be
dismissed for every galaxy model. Here it is possible to show that
when the parameter $k_1$ is in the superior limit of stable curves,
i.e., $k_1\sim 10^{-5}$, the curves become non-planar. It is showed
in Fig. \ref{fig:nonplanar}.

\section{Concluding remarks}\label{sec:conclusions}
Here we presented stable rotation curves of a configuration living
in a $6D$ universe and the comparison with some potentials present
in literature. Initially from a configuration whose density profile
coincides with the Newtonian potential for spiral galaxies our model
was constructed and a Miyamoto-Nagai ansatz was used to solve
Einstein equations. The stable rotation curves of such system were
computed and, without fitting techniques, we recovered with fidelity
the observational data.

The present semi-phenomenological approach points to the evidence
that a universe endowed with extra dimensions could explain the
``missing mass'' problem without a dark matter particle, at least
for the case of rotation curves of galaxies, following the same
results obtained in \cite{coimbra1}. As explained in \cite{coimbra1}
and also in \cite{coimbra2}, although there is no fundamental theory
presented, Eqs. (\ref{lagrange})--(\ref{ydot}) allow one to connect
our model with some universal extra dimension like theory
\cite{appel}, although here we relax about compactification.


Extra dimensions arise here three main integration new constants
($k_1$, $C_x$ and $C_y$), that are used as parameters of the model.
The main goal is to find stable parameters using a geodesic
perturbative approach. Thus we find that $C_x$ and $C_y$ should
range as the conventional thin disk calculation obtained in
\cite{coimbra2}, and $k_1$ is a very tiny parameter ranging from
$10^{-7}-10^{-5}$. Miyamoto-Nagai configurations also introduce the
parameter $b/a$ and for disky galaxies we should use $b/a\sim0.01$.

For the actual solution of $k(R,z)$ we have one of the two extra
dimensions becoming large and the other very small. It is well known
that the important part of the galaxy rotation curve for spirals
(our case) occurs for $0<R<10 R_d$, as can be seen in \cite{binney},
and where $R_d$ is the half luminosity radius of the disk. Thus with
no loss of generality, in the region of interest we do $k = k_1
(z\pm R)$. As explained in section IV, the stability of the model is
guaranteed for $k_1$ very small (here $10^{-5}-10^{-7}$ for $R$
calculated in kpc). When $R>10 R_d$ we have only $k$ greater than
one when $R \sim 10^7$ kpc, a value as great as the radius of the
universe. Thus the smallness of $k_1$ guarantee a approximated
asymptotically flat spacetime in scales of a galaxy.

Our model also, inside the range where curves are stable, reproduces
the behavior of non flat galaxy rotation curves. The incredibly
amount of theories that either imply or assume the existence of an
observational scenario in which rotation curves of spirals are
asymptotically flat, is clearly in contradiction with observational
evidence. Thus, in this aspect, the present model can explain both
planar and non-planar rotation curves.

Here obviously there is no claim to replace the dark matter paradigm
since the usual models of dark matter have been a relatively
successful black box, fitting very well with many theoretical and
observational issues: not only rotation curves and lensing, but also
N--body numerical simulations, structure formation and the analysis
of anisotropies of the CMB. To replace this black box by another one
is not a simple matter, and requires its testing in a wide scope of
theoretical and observational areas.

Also one can interpret the model as part os some theory of universal
extra dimensions. The phenomenology presented has no pretension to
rise aspects of some fundamental theory based on quantum fields. As
we relaxed about compactification and worked on some effects on
large scales (i.e. galaxy scales), it is not in the scope of the
present paper to argue about constraints from particle experiments.


\acknowledgments The authors are very grateful to I.T. Pedron for
important discussions about the main lines of the paper. The work of
C.H. C.-A. is supported by PDEE/CAPES Programme under Grant No.
3874-07-9 and P.S.L. thanks CNPq and FAPESP for partial financial
support. Note: P.S.L. passed away in June, 9th 2011. This a
posthumous publication.

\appendix
\section{Motion equations for a test particle in gravity with extra dimensions}

The Einstein--Hilbert gravitational action with more dimensions is
given by

\begin{equation}\label{action}
S=\int \mathrm{d}^4x\mathrm{d}^n y
\sqrt{-^{(4+n)}g}~(^{(4+n)}R+\mathcal{L}_M),
\end{equation}

\noi This leads to the field equations \bege\label{einstein}
~^{(4+n)}G_{AB}= ~^{(4+n)}T_{AB}, \enge \noi where $A,B = 0, 1, ...,
4+n-1$, $y$ are the extradimensions and the indices $(4+n)$ tells
about the multidimensional nature of the action. From now,
$~^{(4+n)}G_{AB}$ and $~^{(4+n)}T_{AB}$ will be simply called
$G_{AB}$ and $T_{AB}$ (the same for the curvature tensor and scalar,
and as well for the metric).

The most general metric for the space-time given above is

\bege\label{metric1} g(x^\alpha)=\begin{pmatrix}
g_{\alpha\beta}&|&g_{\alpha b}\\
-~-~-&  &-~-~-\\
g_{a\beta}&|&g_{ab}
\end{pmatrix},
\enge

\noi where $\alpha,\beta=(0,..,3)$ and $a,b=(4,..,n)$, for any
integer $n \geqslant 4$, and we are considering the conventional
treatment to do the metric a function only of $3+1$ coordinates.
This metric, as written above, contains the explicit terms meaning
the $3+1$ universe and also the $n$ terms plus crossed components.
In fact, Eq. (\ref{metric1}) can be as well rewritten, for
convenience, in a different way as \bege \label{metric2} g_{AB}=
g_{\alpha\beta}\delta^\alpha_A\delta^\beta_B +
g_{ab}\delta^a_A\delta^b_B+g_{\alpha
b}\delta^\alpha_A\delta^b_B+g_{a\beta}\delta^a_{A}\delta^\beta_B,\enge
\noi for $A,B=(0,..,3+n)$ and where $\delta^i_j$ are the
conventional Kronecker symbols.

Considering the particular case where the metric is diagonal, it is
possible to find the derivatives and consequently the curvature
terms (see \cite{coimbra_accomp} for the explicit calculation). The
equations of motion for such system are calculated as

\begin{equation}\label{motion}
\ddot{x}^\mu + \left\{_{\alpha\beta}^{~\mu}\right\}
\dot{x}^\alpha\dot{x}^\beta =
\frac{1}{2}g_{ab,\gamma}g^{\mu\gamma}N_c g^{ac} N_d g^{bd},
\end{equation}

\noi where $N_a$ means a vector that contains, e.g., the parameters
$C_x$ and $C_y$ (for the case of $6D$). The metric elements should
be calculated by a new Poisson equation (plus boundary conditions
and initial values) that arises from the new terms in Einstein
equations. The Myiamoto-Nagai example in the present paper is a
particular case for such approach. All the discussion about those
terms are presented at \cite{coimbra_accomp}.

Inside the disk galaxy, extra dimensions affect gravity from an
effective potential calculated as

\begin{equation}
\Phi = \phi + C \cosh (k + \delta),
\end{equation}

\noi where $\phi$ is the potential that comes from $4D$, $C$ and $\delta$
are constants to be calculated and $k$ is the function associated
with extra dimensions inside the metric.


\begin{thebibliography}{99}

\bibitem{oort} J. Oort J, Bull. Astron. Inst. Neth. {\bf 6}, 249 (1932); {\it ibid.} {\bf 15}, 45 (1960); F. Zwicky, Helv. Phys. Acta {\bf 6}, 110 (1933); Smith S, Astrophys. J. {\bf 83}, 23 (1936).

\bibitem{sofuerubin} Y. Sofue and V. Rubin, Ann. Rev. Astr. Astrophys. {\bf 39}, 137 (2001).

\bibitem{zwicky} F. Zwicky, Astrophys. J. {\bf 86}, 217 (1937).

\bibitem{fort} B. Fort and Y. Mellier, Astron. Astrophys. Rev. {\bf 5}, 239 (1994); Mellier Y, Ann. Rev. Astron. Astrophys. {\bf 37}, 127 (1999).

\bibitem{coimbra1} C. H. Coimbra-Araújo and P. S. Letelier, Phys.
Rev. D {\bf 76}, 043522 (2007).

\bibitem{rotation} Y. Sofue {\it et al.}, Astrophys. J. {\bf 523}, 136 (1999);
V. Rubin, Int. Astron. Un. Symp. {\bf 117}, 66 (1987); O. Garrido {\it
et al.}, Mon. Not. R. Astron. {\bf 349}, 225 (2004); Y. Sofue {\it
et al.}, Pac. Astr. Soc. J. {\bf 55}, 59 (2003).

\bibitem{steigman} Steigman; B. Moore, Nature {\bf 370}, 629
(1994).

\bibitem{isotropic} M. Miyamoto and N. Nagai, Publications of the Astronomical Society of Japan {\bf 27}, 533
(1975).

(1983).


\bibitem{coimbra_accomp} C. H. Coimbra-Araújo and P. S. Letelier,
{\it Gravity with extra dimensions and dark matter interpretation :
a simple GR approach and cosmological consequences}, in preparation.


\bibitem{vogt} D. Vogt and P. S. Letelier, Mon. Not. R. Astron. {\bf 363}, 268 (2005).

\bibitem{coimbra2} C. H. Coimbra-Araújo and P. S. Letelier,
Classical and Quantum Gravity {\bf 25}, 015001 (2008).

\bibitem{begeman} K. G. Begeman, Astron. \& Astrophys. {\bf 223}, 47 (1989).

\bibitem{navarro} J. F. Navarro, C. S. Frenk and S. D. M. White, Astrophys. J. {\bf 490}, 493 (1997).

\bibitem{courteau} S. Courteau, Astron. J. {\bf 114}, 2402 (1997).

\bibitem{binney} J. Binney and S. Tremaine, {\it Galactic Dynamics} (Princenton, Princenton University Press, 1987).

\bibitem{appel} T. Appelquist, H. C. Cheng and B. A. Dobrescu, Phys. Rev. D{\bf 64}, 035002 (2001).

\bibitem{martins} C. Frigerio Martins, {\it The distribution of the dark matter in galaxies as the imprint of its Nature}, Ph.D. Thesis, SISSA (2008).

\bibitem{coimbra3} C. H. Coimbra-Araújo and P. S. Letelier,
\emph{Proceedings of IAU Symposium n. 245}, edited by V. Karas
(Cambridge Un. Press, Cambridge, 2007) pp 239-240.




\end{thebibliography}
\end{document}